\documentclass[preprint,superscriptaddress,showpacs,tightenlines]{revtex4}
\usepackage{amssymb}
\usepackage{amsmath}
\usepackage{graphicx,bm}

\input{tcilatex}

\begin{document}

\title{Method to determine defect positions below a metal \\
surface by STM }
\author{Ye.S. Avotina}
\affiliation{B.I. Verkin Institute for Low Temperature Physics and Engineering, National
Academy of Sciences of Ukraine, 47, Lenin Ave., 61103, Kharkov,Ukraine.}
\affiliation{Kamerlingh Onnes Laboratorium, Universiteit Leiden, Postbus 9504, 2300
Leiden, The Netherlands.}
\author{Yu.A. Kolesnichenko}
\affiliation{B.I. Verkin Institute for Low Temperature Physics and Engineering, National
Academy of Sciences of Ukraine, 47, Lenin Ave., 61103, Kharkov,Ukraine.}
\author{A.N. Omelyanchouk}
\affiliation{B.I. Verkin Institute for Low Temperature Physics and Engineering, National
Academy of Sciences of Ukraine, 47, Lenin Ave., 61103, Kharkov,Ukraine.}
\author{A.F. Otte}
\affiliation{Kamerlingh Onnes Laboratorium, Universiteit Leiden, Postbus 9504, 2300
Leiden, The Netherlands.}
\author{J.M. van Ruitenbeek}
\affiliation{Kamerlingh Onnes Laboratorium, Universiteit Leiden, Postbus 9504, 2300
Leiden, The Netherlands.}
\maketitle

\section{Introduction.}

In the two decades following its invention, scanning tunnelling microscopy
(STM) has proved to be a valuable tool for investigating surfaces on an
atomic scale. More recently, several experiments show a growing interest in
the study of structures that are situated in the bulk below the surface in
both semiconductors and metals. Whereas in the former (i.e. semiconductor)
case the absence of effective screening allows dopants down to the third
subsurface layer to be viewed directly as apparent topographic features,\cite%
{Ebert} the situation for metals turns out to be somewhat more complicated.
One method that has been suggested for imaging structures buried in metal
involves several surface study techniques to be employed simultaneously in
combination with STM\cite{Heinze}. However, although this experiment has
lead to successful identification of subsurface defects, it cannot be used
as a tool for probing the exact depth. Also, its employability is limited to
certain specific alloys only. A more successful approach however seems to be
by probing standing electron waves\cite{Schmid,Hongbin}. The groundwork of
these experiments is as described in Ref.\cite{Crommie}, where Cu(111)
surface states form a two-dimensional nearly free electron gas. When
scattered from step edges or adatoms, these states then form standing waves
which can be probed by scanning tunnelling spectroscopy (STS).

Although it has been proposed to utilize these surface states for imaging
subsurface impurities\cite{Crampin}, the exponential decay of the wave
function amplitudes into the bulk will limit the effective range to the
topmost layers only. Bulk states however, of which the square falls of\ with
only $r^{2}$, form a good alternative. To demonstrate this, we mention
results that were obtained by bulk state spectroscopy on relatively large
structures such as Ar bubbles submerged in Al, \cite{Schmid} and Si(111)
step edges buried under a thin film of Pb \cite{Hongbin}. In these
experiments, bulk electrons are found to be confined in a vertical quantum
well between the surface and the top plane of the object of interest.

In this paper we show that the investigation of the nonlinear conductance of
a point contact placed on a metal surface makes it possible to determine the
position of point-like defects such as the vacancies or foreign atoms inside
the metal in the vicinity of the contact. We consider theoretical models
both for the cases of a tunnel point contact and for a ballistic quantum
contact. We look for conductance oscillations caused by interference of
electrons that are transmitted directly, and electrons that are first
backscattered elastically by the defect and again scattered forward by the
contact (i.e. the tip-sample junction), much in the same way as was
described for atomic point contacts in Refs. \cite{Ludoph,Untiedt,Kempen}
The effect of such quantum interference on the nonlinear conductance of
quantum wires was theoretically analyzed in Refs.\cite%
{Namir,Avotina,Avotina1}, but the point contact geometry was not studied yet.

The organization of this paper is as follows. In Sec.~II we consider a
tunnel junction in the limit of a high potential barrier. The interaction of
the transmitted electrons with a single impurity near the junction is taken
into account by perturbation theory with the electron-impurity interaction
as the small parameter. A general analytical expression for the voltage
dependence of conductance, $G\left( V\right) $, is obtained. It defines $%
G\left( V\right) $ in terms of the contact diameter, the distance between
contact and the impurity and the parameters that characterize the metal, and
the transmission of the tunnel junction. In Sec.~III the conductance of a
ballistic quantum contact of adiabatic shape is analyzed. In absence of a
barrier inside the contact electrons can still be reflected from it due to
the variation of the confining potential. The influence on electron
scattering by an explicit barrier potential in the center of the contact is
also discussed. As in Sec.~II assuming the electron-impurity interaction to
be small we derive an expression for $G\left( V\right) $ and its dependence
on the position of the defect. In Sec.~IV we conclude by discussing the
possibilities for experimental exploitation of the conductance fluctuations
for sub-surface imaging as well as the technical difficulties involved.

\section{\protect\bigskip Tunnel point-contact.}

\begin{figure}[t]
\includegraphics[width=0.7\linewidth]{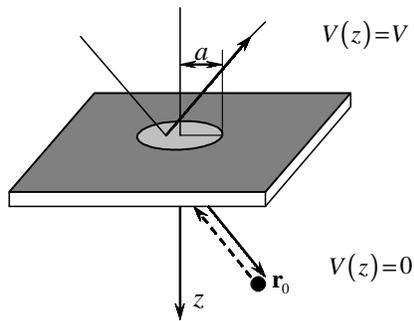}
\caption{Model of a tunnel junction contact as an orifice in an interface
that is nontransparent for electrons except for a circular hole, where
tunnelling is allowed. Trajectories are shown schematically for electrons
that are reflected from, or transmitted through the contact and then
reflected from a defect.}
\label{fig.orifices}
\end{figure}

\bigskip Let us consider as a first model of our system a nontransparent
interface located at $z=0$ between two metal half-spaces, in which there is
an orifice (contact), as illustrated in Fig.~\ref{fig.orifices}. The
potential barrier in the plane $z=0$ is taken to be a $\delta -$function: 
\begin{equation}
U\left( \mathbf{r}\right) =Uf\left( \mathbf{\rho }\right) \delta \left(
z\right) ,
\end{equation}
where $\mathbf{\rho }=\left( x,y\right) $ is a two dimensional vector. The
function $f\left( \mathbf{\rho }\right) \rightarrow \infty $ in all points
of the plane $z=0$ except in the contact, where $f\left( \mathbf{\rho }%
\right) =1$. At a point $\mathbf{r}=\mathbf{r}_{0}$ in vicinity of the
interface, in the half-space $z>0$, a point-like defect is placed, see Fig.~%
\ref{fig.orifices}. The electron interaction with the defect is described by
the potential 
\begin{equation}
g\left( \mathbf{r}\right) =g\delta \left( \mathbf{r}-\mathbf{r}_{0}\right) ,
\end{equation}
where $g$ is the constant of the electron-impurity interaction. In this
section we consider tunnel junctions and assume that the transmission
probability of electrons through the orifice is small. In that case the
applied voltage drops entirely over the barrier and we choose the electric
potential as a step function $V\left( z\right) =V\,\Theta \left( -z\right) ,$
and take $eV>0.$

\bigskip The electrical current $I\left( V\right) $ can be evaluated\cite%
{ISh} from the electron wave functions of the system, $\psi _{\mathbf{k}}$, 
\begin{equation}
I\left( V\right) =\frac{e\hbar }{4\pi ^{3}m^{\ast }}\int d\mathbf{k}\int d%
\mathbf{S}\ \mathrm{Im}\left( \psi _{\mathbf{k}}^{\ast }\nabla \psi _{%
\mathbf{k}}\right) \Theta \left( k_{z}\right) \left[ n_{F}\left( \varepsilon
_{\mathbf{\ k}}\right) -n_{F}\left( \varepsilon _{\mathbf{k}}+eV\right) %
\right] .  \label{current}
\end{equation}
Here, $\varepsilon _{\mathbf{k}}=\hslash ^{2}\mathbf{k}^{2}/2m^{\ast }$ is
the electron energy, $\mathbf{k}$ is the electron wave vector and $m^{\ast }$
is an effective mass of electron; $n_{F}\left( \varepsilon _{\mathbf{k}%
}\right) $ is the Fermi distribution function. The real space integration is
performed over a surface overlapping the contacts in the region $z>0.$ At
low temperatures the tunnel current is due to those electrons in the
half-space $z<0$ having an energy between the Fermi energy, $\varepsilon
_{F},$ and $\varepsilon _{F}+eV$, because on the other side of the barrier
only states with $\varepsilon _{\mathbf{k}}\geqslant \varepsilon _{F}$ are
available.

The wave function $\psi _{\mathbf{k}}$ satisfies the Schr\"{o}dinger
equation \ \ \ 
\begin{equation}
\triangledown ^{2}\psi _{\mathbf{k}}\left( \mathbf{\rho },z\right) +\frac{%
2m^{\ast }}{\hslash ^{2}}\left[ \varepsilon _{\mathbf{k}}-U\left( \mathbf{r}%
\right) -g\left( \mathbf{r}\right) -eV\left( z\right) \right] \psi _{\mathbf{%
k}}\left( \mathbf{\rho },z\right) =0  \label{Schrod}
\end{equation}
where the wave vector $\mathbf{k=}\left( \mathbf{\varkappa ,}k_{z}\right) $
has components $\mathbf{\varkappa }$ and $k_{z}$ parallel and perpendicular
to interface, respectively. As shown in Ref.\cite{KMO}, Eq.~(\ref{Schrod})
can be solved for arbitrary form of the function $f\left( \mathbf{\rho }%
\right) $ in the limit $1/U\rightarrow 0.$ The wave function $\psi _{\mathbf{%
k}}\left( \mathbf{\rho },z\right) $ for $k_{z}>0$ in the main approximation
of the small parameter $\sim 1/U$ takes the form: 
\begin{eqnarray}
\psi _{\widetilde{\mathbf{k}}}\left( \mathbf{\rho },z\right) &=&e^{i\mathbf{%
\varkappa \rho }}\left( e^{i\widetilde{k}z}-e^{-i\widetilde{k}z}\right) +%
\frac{1}{U}\varphi _{\widetilde{\mathbf{k}}}^{\left( -\right) }\left( 
\mathbf{\rho },z\right) \qquad (z<0), \\
\psi _{\mathbf{k}}\left( \mathbf{\rho },z\right) &=&\frac{1}{U}\varphi _{%
\mathbf{k}}^{\left( +\right) }\left( \mathbf{\rho },z\right) \quad \qquad
\qquad \qquad \qquad \qquad (z>0),  \label{psiz>0}
\end{eqnarray}
where $\widetilde{\mathbf{k}}\mathbf{=}\left( \mathbf{\varkappa ,-}%
\widetilde{k}\right) ,$ $\widetilde{k}=\sqrt{k_{z}^{2}+2meV/\hbar ^{2}}$.
The function $\psi _{\mathbf{k}}\left( \mathbf{\rho },z\right) $ satisfies
the conditions of continuity and the condition of the jump of its derivative
at the boundary $z=0.$ At large $U$ these conditions are reduced to 
\begin{gather}
\varphi _{\widetilde{\mathbf{k}}}^{\left( -\right) }\left( \mathbf{\rho }%
,0\right) =\varphi _{\mathbf{k}}^{\left( +\right) }\left( \mathbf{\rho }%
,0\right) ;  \label{bound1} \\
i\widetilde{k}=\frac{m^{\ast }}{\hbar ^{2}}f\left( \mathbf{\rho }\right)
\varphi _{\mathbf{k}}^{\left( +\right) }\left( \mathbf{\rho },0\right) .
\label{gran_cond}
\end{gather}
In the absence of the defect $\left( g=0\right) $ the wave function was
obtained in Ref.\cite{KMO}, 
\begin{equation}
\varphi _{0\mathbf{k}}^{\left( +\right) }\left( \mathbf{\rho },z\right) =-%
\frac{i\hslash ^{2}\widetilde{k}}{2\pi m^{\ast }}\int\limits_{-\infty
}^{\infty }d\mathbf{\varkappa }^{\prime }F\left( \mathbf{\varkappa }-\mathbf{%
\varkappa }^{\prime }\right) e^{i\mathbf{\varkappa }^{\prime }\mathbf{\rho }%
+ik_{z}^{\prime }z},  \label{phi+0}
\end{equation}
where 
\begin{equation}
F\left( \mathbf{\varkappa }-\mathbf{\varkappa }^{\prime }\right)
=\int\limits_{-\infty }^{\infty }d\mathbf{\rho }\frac{e^{i\left( \mathbf{%
\varkappa }-\mathbf{\varkappa }^{\prime }\right) \mathbf{\rho }}}{f\left( 
\mathbf{\rho }\right) },
\end{equation}
and $k_{z}^{\prime }=\sqrt{\mathbf{\varkappa }^{2}+k_{z}^{2}-\mathbf{%
\varkappa }^{\prime 2}}.$ For a circular contact of a radius $a$, defined by 
$f\left( \left| \mathbf{\rho }\right| \leqslant a\right) =1$ and $f\left(
\left| \mathbf{\rho }\right| >a\right) \rightarrow \infty $, the function $%
F\left( \mathbf{\varkappa }-\mathbf{\varkappa }^{\prime }\right) $ takes the
form 
\begin{equation}
F\left( \mathbf{\varkappa }-\mathbf{\varkappa }^{\prime }\right) =\frac{%
2aJ_{1}\left( \left| \mathbf{\varkappa }-\mathbf{\varkappa }^{\prime
}\right| a\right) }{\left| \mathbf{\varkappa }-\mathbf{\varkappa }^{\prime
}\right| }.
\end{equation}

\begin{figure}[t]
\includegraphics[width=0.7\linewidth]{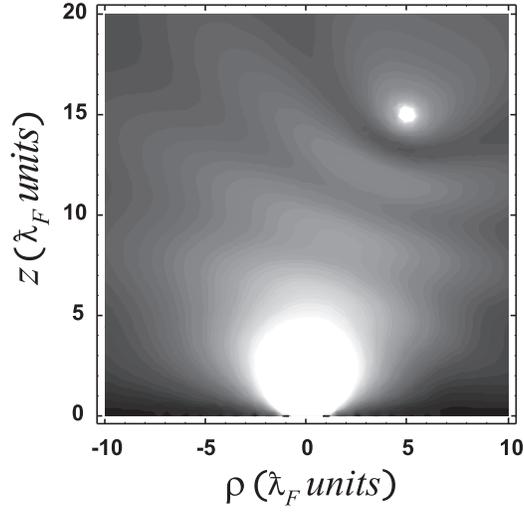}
\caption{Modulus of the wave function in the vicinity of a tunnelling
point-contact in a plane perpendicular to the contact axis, having an
impurity at ($\protect\rho _{0}=5,$ $z_{0}=15$). The incident wave has a
wave vector normal to the point contact $a=\protect\lambda _{F}/4\protect\pi 
$.}
\label{fig.gray-scale}
\end{figure}

In order to introduce the effect of the impurity we solve the Schr{\"{o}}%
dinger equation for the Fourier components $\Phi _{\mathbf{k}}\left( \mathbf{%
\varkappa },z\right) $ of the function $\varphi _{\mathbf{k}}^{\left(
+\right) }\left( \mathbf{\rho },z\right) $ 
\begin{equation}
\varphi _{\mathbf{k}}^{\left( +\right) }\left( \mathbf{\rho },z\right)
=\int\limits_{-\infty }^{\infty }d\mathbf{\varkappa }e^{i\mathbf{\varkappa
\rho }}\Phi _{\mathbf{k}}\left( \mathbf{\varkappa },z\right) .
\end{equation}
For $z>0$ this equation takes the form 
\begin{gather}
-\mathbf{\varkappa }^{\prime 2}\Phi _{\mathbf{k}}\left( \mathbf{\varkappa }%
^{\prime },z\right) +\frac{\partial ^{2}\Phi _{\mathbf{k}}\left( \mathbf{\
\varkappa }^{\prime },z\right) }{\partial z^{2}}+\frac{2m^{\ast }}{\hslash
^{2}}\left[ \varepsilon _{\mathbf{k}}\Phi _{\mathbf{k}}\left( \mathbf{\
\varkappa }^{\prime },z\right) -\right.  \label{Schrod_Fourier} \\
\left. g\delta \left( z-z_{0}\right) e^{-i\mathbf{\varkappa }^{\prime }%
\mathbf{\rho }_{0}}\varphi _{\mathbf{k}}^{\left( +\right) }\left( \mathbf{\
\rho }_{0},z_{0}\right) \right] =0  \notag
\end{gather}
Integrating Eq.~(\ref{Schrod_Fourier}) near the point $z=z_{0}$ we obtain
the effective boundary condition: 
\begin{equation}
\frac{\partial }{\partial z}\Phi _{\mathbf{k}}\left( \mathbf{\varkappa }%
^{\prime },z=z_{0}+0\right) -\frac{\partial }{\partial z}\Phi _{\mathbf{k}%
}\left( \mathbf{\varkappa }^{\prime },z=z_{0}-0\right) =\frac{2m^{\ast }g}{%
\hslash ^{2}}e^{-i\mathbf{\varkappa }^{\prime }\mathbf{\rho }_{0}}\varphi _{%
\mathbf{k}}^{\left( +\right) }\left( \mathbf{\rho }_{0},z_{0}\right)
\label{Jump_Fourier}
\end{equation}
To proceed with further calculations we assume that the electron-impurity
interaction constant $g$ is small and use perturbation theory. In this
approximation we replace $\varphi _{\mathbf{k}}^{\left( +\right) }$ by $%
\varphi _{0\mathbf{k}}^{\left( +\right) }$ (\ref{phi+0}) in the right hand
side of Eq.~(\ref{Jump_Fourier}). Solving the Schr\"{o}dinger equation (\ref%
{Schrod_Fourier}) with the boundary conditions (\ref{bound1}), (\ref%
{gran_cond}), (\ref{Jump_Fourier}), and the condition of continuity of the
function $\Phi _{\mathbf{k}}\left( \mathbf{\varkappa },z\right) $ at $%
z=z_{0},$ we obtain in the region $z>z_{0}$ 
\begin{equation}
\Phi _{\mathbf{k}}\left( \mathbf{\varkappa }^{\prime },z\right) =t_{\mathbf{k%
}}\left( \mathbf{\varkappa }^{\prime }\right) e^{ik_{z}^{\prime }z},
\label{Fi_t}
\end{equation}
where 
\begin{equation}
t_{\mathbf{k}}\left( \mathbf{\varkappa }^{\prime }\right) =-\frac{i\hslash
^{2}\widetilde{k}}{2\pi m^{\ast }}F\left( \mathbf{\varkappa }-\mathbf{%
\varkappa }^{\prime }\right) -\frac{2m^{\ast }g}{k_{z}^{\prime }\hslash ^{2}}%
\varphi _{0\mathbf{k}}^{\left( +\right) }\left( \mathbf{\rho }%
_{0},z_{0}\right) e^{-i\mathbf{\varkappa }^{\prime }\mathbf{\rho }_{0}}\sin
\left( k_{z}^{\prime }z_{0}\right) .  \label{t}
\end{equation}
Using Eq.~(\ref{Schrod_Fourier}) we find the wave function 
\begin{gather}
\varphi _{\mathbf{k}}^{\left( +\right) }\left( \mathbf{\rho },z\right)
=\varphi _{0\mathbf{k}}^{\left( +\right) }\left( \mathbf{\rho },z\right) -
\label{Funct_wave} \\
\frac{4\pi m^{\ast }g}{\hslash ^{2}}\varphi _{0\mathbf{k}}^{\left( +\right)
}\left( \mathbf{\rho }_{0},z_{0}\right) \int\limits_{0}^{\infty }d\varkappa
^{\prime }\varkappa ^{\prime }e^{ik_{z}^{\prime }z_{1}}\frac{\sin \left(
k_{z}^{\prime }z_{2}\right) }{k_{z}^{\prime }}J_{0}\left( \varkappa ^{\prime
}\left| \mathbf{\rho }-\mathbf{\rho }_{0}\right| \right) ,  \notag
\end{gather}
where $z_{1}=z,$ $z_{2}=z_{0},$ when $z>z_{0},$ and $z_{1}=z_{0},$ $z_{2}=z$
when $0<z<z_{0}$. The modulus of the wave function (\ref{Funct_wave}) in a
plane through the impurity and normal to the circular contact is illustrated
in Fig.~\ref{fig.gray-scale}, for an incident wave vector normal to the
interface ($k_{z}=k_{F}=\frac{1}{\hbar }\sqrt{2m^{\ast }\varepsilon _{F}}$).
One recognizes a interference pattern of partial waves reflected at the
impurity with those emanating from the contact. These contain the
information that we hope to extract.

Substituting wave function (\ref{psiz>0}) into Eq.~(\ref{current}) and
taking into account Eq.~(\ref{Funct_wave}), we calculate the current-voltage
characteristics $I\left( V\right) $. After integration over all directions
of the wave vector $\mathbf{k}$ and integration over the space coordinate $%
\mathbf{\rho }$ in a plane $z=const $ ($z>z_{0}$), retaining only terms to
first order in $g$ (i.e. ignoring multiple scattering at the impurity site),
the current is given by 
\begin{gather}
I\left( V\right) =\frac{2e\hbar ^{5}}{2\pi m^{\ast 3}U^{2}}%
\int\limits_{0}^{\infty }dkk^{3}\iint \frac{d\mathbf{\rho }_{1}d\mathbf{\rho 
}_{2}}{f\left( \mathbf{\rho }_{1}\right) f\left( \mathbf{\rho }_{2}\right) }%
\left[ n_{F}\left( \varepsilon _{\mathbf{k}}\right) -n_{F}\left( \varepsilon
_{\mathbf{k}}+eV\right) \right]  \label{I(V)} \\
\left[ \frac{A^{2}\left( k\rho \right) }{\rho ^{4}}-2\pi \frac{m^{\ast }gk}{%
\hbar ^{2}}\frac{A\left( k\lambda _{1}\right) A\left( k\lambda _{2}\right) }{%
\lambda _{2}^{2}\lambda _{1}^{2}}z_{0}^{2}\left( \frac{A\left( k\rho \right) 
}{\rho ^{2}}+\frac{2m^{\ast }eV}{\hbar ^{2}}\frac{\sin \left( k\rho \right) 
}{k\rho }\right) \right] ,  \notag
\end{gather}
where $\rho =\left| \mathbf{\rho }_{1}-\mathbf{\rho }_{2}\right| ,$ $\lambda
_{1}=\sqrt{z_{0}^{2}+\left| \mathbf{\rho }_{0}-\mathbf{\rho }_{1}\right| ^{2}%
},$ $\lambda _{2}=\sqrt{z_{0}^{2}+\left| \mathbf{\rho }_{0}-\mathbf{\rho }%
_{2}\right| ^{2}},$ and 
\begin{equation}
A\left( x\right) =\frac{\sin x}{x}-\cos x\quad .
\end{equation}
Differentiating Eq.~(\ref{I(V)}) with voltage $V$ and integrating over the
absolute value $k$ of the wave vector, in the limit of low temperatures, $%
T=0 $, we obtain the conductance $G\left( V\right) $ of the system 
\begin{gather}
G\left( V\right) =\frac{e^{2}\widetilde{k}_{F}^{2}\hbar ^{3}}{\pi \left(
m^{\ast }U\right) ^{2}}\iint \frac{d\mathbf{\rho }_{1}d\mathbf{\rho }_{2}}{%
f\left( \mathbf{\rho }_{1}\right) f\left( \mathbf{\rho }_{2}\right) }\left[ 
\frac{A^{2}\left( \widetilde{k}_{F}\rho \right) }{\rho ^{4}}-\right.
\label{G(V)} \\
2\pi \frac{m^{\ast }g\widetilde{k}_{F}}{\hbar ^{2}}\frac{A\left( \widetilde{k%
}_{F}\lambda _{1}\right) A\left( \widetilde{k}_{F}\lambda _{2}\right) }{%
\lambda _{1}^{2}\lambda _{2}^{2}}z_{0}^{2}\left( \frac{A\left( \widetilde{k}%
_{F}\rho \right) }{\rho ^{2}}+\frac{2m^{\ast }eV}{\hbar ^{2}}\frac{\sin
\left( \widetilde{k}_{F}\rho \right) }{\widetilde{k}_{F}\rho }\right) - 
\notag \\
\left. \frac{8\pi }{\widetilde{k}_{F}^{2}}\frac{m^{\ast }g}{\hbar ^{2}}%
z_{0}^{2}\int\limits_{k_{F}}^{\widetilde{k}_{F}}k^{3}dk\frac{\sin \left(
k\rho \right) }{\rho }\frac{A\left( k\lambda _{1}\right) A\left( k\lambda
_{2}\right) }{\lambda _{1}^{2}\lambda _{2}^{2}}\right] ,  \notag
\end{gather}
where $\widetilde{k}_{F}=\sqrt{k_{F}^{2}+2m^{\ast }eV/\hbar ^{2}}$ is the
Fermi wave vector accelerated by the potential difference, and we have
assumed $eV/\varepsilon _{F}<1.$ For $\lambda _{i}$ and $\rho $ much larger
than the Fermi wave length $\lambda _{F}=1/k_{F},$ the function $A$
oscillates with $(eV/\varepsilon _{F})k_{F}\lambda _{i}$ and $%
(eV/\varepsilon _{F})k_{F}\rho $, which results in a fluctuation of the
conductance with applied voltage. The first term is square brackets of Eq.(%
\ref{G(V)}) describes the conductance $G\left( V\right) =G_{c}\left(
V\right) $ in the absence of a defect $\left( g=0\right) $.

For a contact of small diameter $a\ll \lambda _{F}$ Eq.~(\ref{G(V)}) can be
simplified and the conductance is given by 
\begin{figure}[t]
\includegraphics[width=0.7\linewidth]{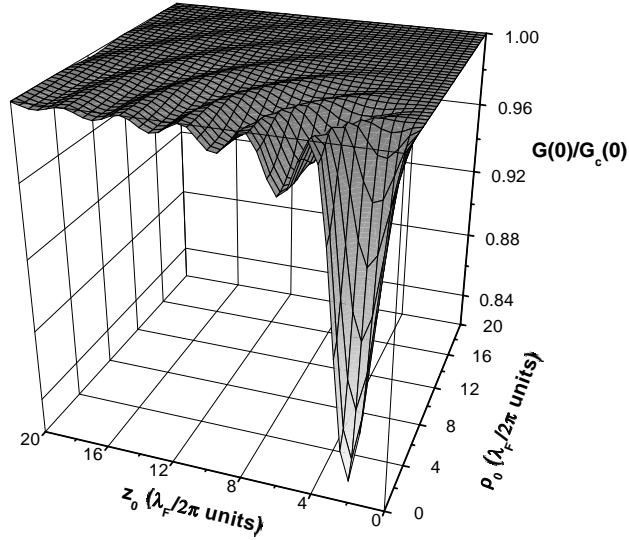} .
\caption{Dependence of the normalized conductance $G/G_{c}$ for a single
tunnel point contact as a function of the position of the defect $\left( 
\protect\rho _{0},z_{0}\right) $, contact radius is $a=\protect\lambda _{F}/4%
\protect\pi $ . }
\label{fig.G_vs_Z0_and_rho}
\end{figure}
\begin{eqnarray}
G\left( V\right) &=&G_{c}\left( V\right) \left\{ 1-\frac{6\pi m^{\ast }g}{%
\hbar ^{2}}\frac{A^{2}\left( \widetilde{k}_{F}\gamma \right) }{\widetilde{k}%
_{F}\gamma ^{4}}z_{0}^{2}\left( 1+\frac{6m^{\ast }eV}{\hbar ^{2}\widetilde{k}%
_{F}^{2}}\right) -\right.  \label{G_one} \\
&&\left. \frac{72\pi gm^{\ast }}{\hbar ^{2}\widetilde{k}_{F}^{7}}\frac{%
z_{0}^{2}}{\gamma ^{4}}\int\limits_{k_{F}}^{\widetilde{k}_{F}}dkk^{4}A^{2}%
\left( k\gamma \right) \right\} ,  \notag
\end{eqnarray}
where $\gamma =\sqrt{z_{0}^{2}+\left| \rho _{0}\right| ^{2}}$ is the
distance between the contact and the defect. The conductance $G_{c}\left(
V\right) $ of the tunnel junction of small cross section $S=\pi a^{2}\ll
\lambda _{F}^{2}$ is given by 
\begin{equation}
G_{c}\left( V\right) =\frac{4\pi e^{2}\varepsilon _{F}^{2}\widetilde{k}%
_{F}^{2}a^{4}}{9\hbar U^{2}}=\frac{\pi ^{2}}{9}T_{b}\left( \widetilde{k}%
_{F}\right) \frac{2e^{2}}{h}\left( \widetilde{k}_{F}a\right) ^{4},
\label{G_c}
\end{equation}
for small transmission coefficient $T_{b}\left( k\right) =\hbar ^{4}k^{2}/m{%
^{\ast 2}}U{^{2}}\ll 1.$

\begin{figure}[t]
\includegraphics[width=0.7\linewidth]{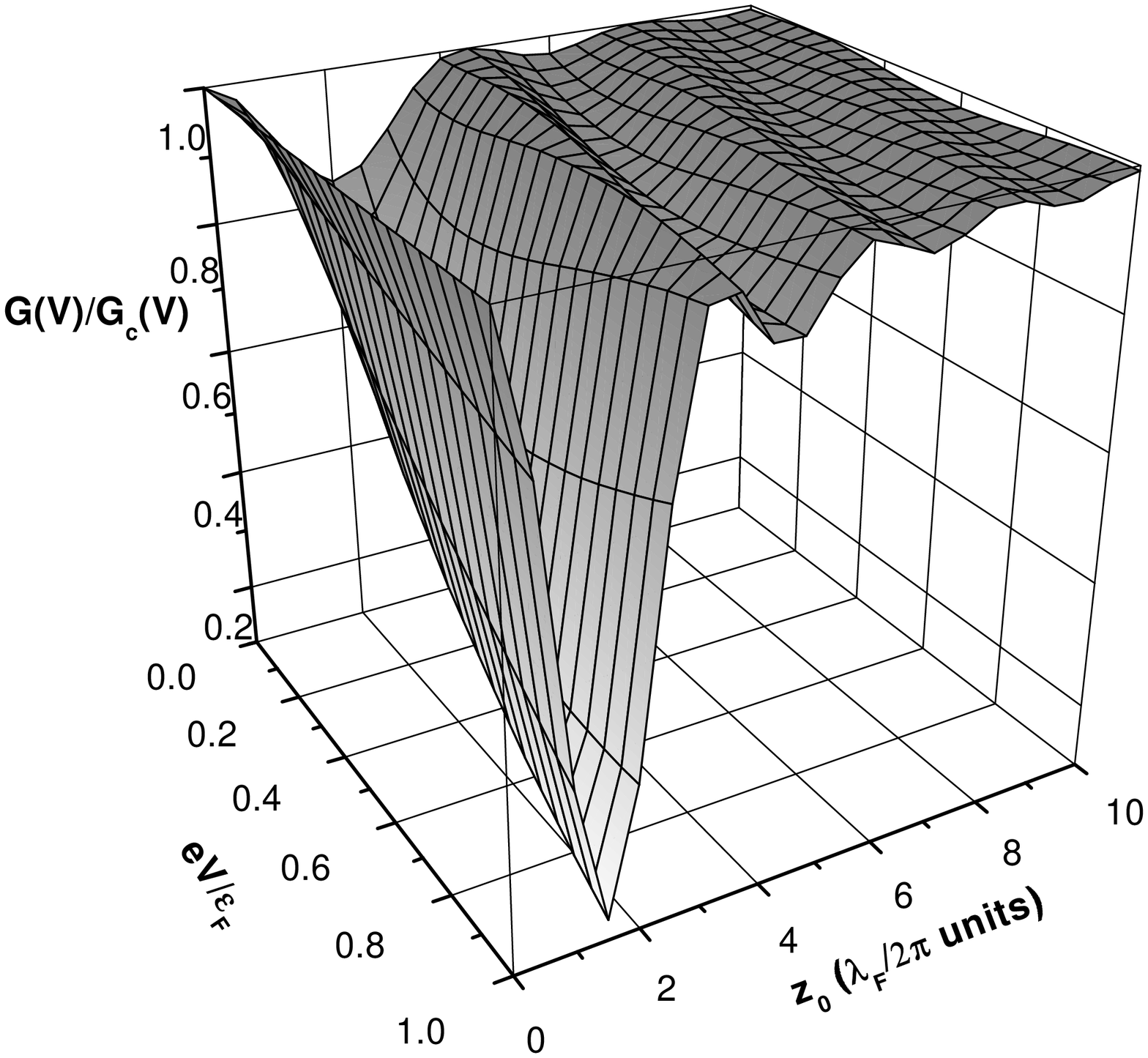}
\caption{Voltage dependence of the normalized conductance $G\left( V\right)
/G_{c}(V)$ of a tunnel point contact for $\protect\rho _{0}=0$ and as a
function of the depth $z_{0}$ of the defect under metal surface, contact
radius is $a=\protect\lambda _{F}/4\protect\pi $. }
\label{fig.G_vs_Z0_and_V}
\end{figure}

In numerical calculations we use a value for the dimensionless parameter $%
2\pi m^{\ast }gk_{F}/\hbar ^{2}=0.3$ to characterize the strength of the
defect scattering. Fig.~\ref{fig.G_vs_Z0_and_rho} shows a plot of the
dependence of the normalized conductance $G\left( 0\right) /G_{c}\left(
0\right) $, Eq.~(\ref{G(V)}), for the contact as a function of the position
of the defect $\left( \rho _{0},z_{0}\right) $ in the limit of low voltage $%
V\rightarrow 0$. We observe a suppression of the conductance that is largest
when the contact is placed directly above the defect and find that $G$ is an
oscillatory function of the defect position. In Fig.~\ref{fig.G_vs_Z0_and_V}
we show the voltage dependence of the normalized conductance $G\left(
V\right) /G_{c}\left( V\right) $, Eq.~(\ref{G(V)}), for $\rho _{0}=0$ and as
a function of the depth $z_{0}$ of the defect under the metal surface.

\FRAME{ftbpF}{4.6449in}{3.2171in}{0in}{}{}{fig3.eps}{\special{language
"Scientific Word";type "GRAPHIC";maintain-aspect-ratio TRUE;display
"USEDEF";valid_file "F";width 4.6449in;height 3.2171in;depth
0in;original-width 11.1301in;original-height 7.689in;cropleft "0";croptop
"1";cropright "1";cropbottom "0";filename 'Fig3.eps';file-properties
"XNPEU";}}

\section{Ballistic contact.}

\begin{figure}[t]
\includegraphics[width=0.7\linewidth]{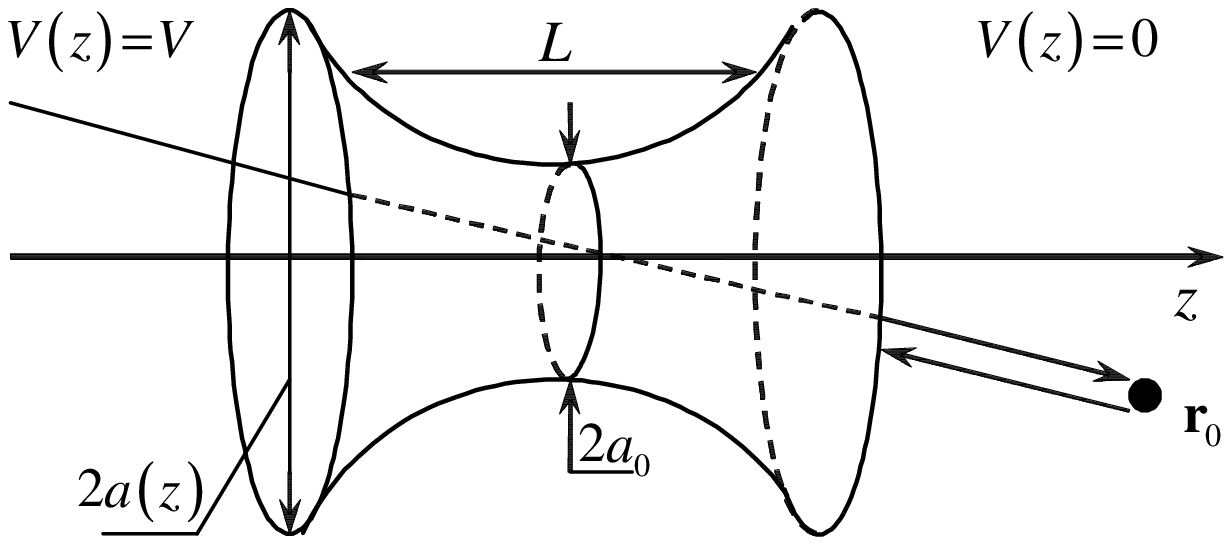}
\caption{Model of a ballistic contact of adiabatic shape having a defect
sitting nearby. The trajectories of electrons that are transmitted through
the contact and reflected from the defect are shown schematically.}
\label{fig.adiabatic_constriction}
\end{figure}

In this section we consider another limit of a junction, a cylindrically
symmetric, ballistic contact of adiabatic shape, Fig.~\ref%
{fig.adiabatic_constriction}. The center of the contact is characterized by
a $\delta $-function potential barrier of amplitude $U.$ In one of the banks
of the contact a single defect is situated at the point $\mathbf{r}%
_{0}=\left( \mathbf{\rho }_{0},z_{0}\right) ,$ in the half-space $z_{0}>0$,
such that the distance $\gamma $ between the center of the contact $\mathbf{r%
}=0$ and the defect is much larger than the characteristic length $L$ of the
constriction (see Fig.~\ref{fig.adiabatic_constriction}). The shape of the
contact is described by the radius as a function of the $z$-coordinate, $%
a\left( z\right) .$ The contact size is given by $a\left( 0\right) =a_{0},$
while $a\left( z\right) \rightarrow \infty $ for $\left| z\right|
\rightarrow \infty $. The adiabatic condition implies that the radius of the
contact $a\left( z\right) $ varies slowly on the scale of the Fermi
wavelength. As a result, the electric potential $V\left( \mathbf{r}\right) $
drops dominantly over the same characteristic length $L$, as can be derived
from the condition of electroneutrality. In the Landauer formalism the exact
distribution of $V\left( \mathbf{r}\right) $ is not important for
determining the conductance of a quantum constriction, which can be
expressed using only the difference of potentials $V$ in the banks far from
the contact. We will consider the effect of quantum interference on the
conductance under conditions $eV/\varepsilon _{F}\ll 1$ and $k_{F}\gamma
(eV/\varepsilon _{F})>1$. Fluctuations of $G\left( V\right) $ results from
the phase shift $\Delta \varphi $\ that the wave function accumulates after
being scattered by the defect and reflected by the contact, $\Delta \varphi $
$\sim (eV/\varepsilon _{F})k_{F}\gamma .$ If $\gamma \gg L,$ the main part
of the electron trajectory is situated in the region where the local
electric potential $V\left( \mathbf{r}\right) $ differs only little from its
value $V=0$ in the bank at $z\rightarrow \infty $, and we neglect this small
variation of the potential. Assuming hard wall boundary conditions, we need
to solve the Schr\"{o}dinger equation, 
\begin{equation}
\triangledown ^{2}\psi _{\mathbf{\alpha }}\left( \mathbf{\rho },z\right) +%
\frac{2m^{\ast }}{\hslash ^{2}}\left[ \varepsilon -g\delta \left( \mathbf{r}-%
\mathbf{r}_{0}\right) -U\delta \left( z\right) \right] \psi _{\mathbf{\alpha 
}}\left( \mathbf{\rho },z\right) =0,  \label{Schred_bal}
\end{equation}
with the boundary conditions 
\begin{equation}
\psi _{\mathbf{\alpha }}\left( \left| \mathbf{\rho }\right| =a(z\right)
;z)=0,
\end{equation}
and $\mathbf{\alpha }$ represents the full set of quantum numbers.

In the adiabatic approximation \cite{Glazman,3D} the ``fast'' transverse and
``slow'' longitudinal variables in Eq.~(\ref{Schred_bal}) can be separated
and the wave function takes on the form 
\begin{equation}
\psi _{\mathbf{\alpha }}\left( \mathbf{\rho },z\right) =\psi _{\bot \beta
}\left( \mathbf{\rho },z\right) \varphi _{\beta \varepsilon }\left( z\right)
,  \label{psi_separ}
\end{equation}
where $\beta =\left( m,n\right) $ is a set of two discrete quantum numbers,
which define the transverse local eigenvalues $\varepsilon _{\beta }\left(
z\right) $ and eigenfunctions $\psi _{\bot \beta }\left( \mathbf{\rho }%
,z\right).$ The function $\psi _{\bot \beta }\left( \mathbf{\rho } ,z\right) 
$ depends on the coordinate $z$ as a local parameter, and its derivatives
with respect to $z$ are small. Therefore Eq.~(\ref{Schred_bal} ) \ can be
separated into two equations, 
\begin{equation}
\triangledown _{\mathbf{\rho }}^{2}\psi _{\bot \beta }\left( \mathbf{\rho }
\right) =\frac{2m^{\ast }}{\hslash ^{2}}\varepsilon _{\beta }\left( a\right)
;  \label{Schred_ro}
\end{equation}
\begin{equation}
\frac{d^{2}\varphi _{\beta \varepsilon }}{dz^{2}}+\frac{2m^{\ast }}{\hslash
^{2}}\left[ \varepsilon -\varepsilon _{\beta }\left( a\right) \right] =0.
\label{Schred_z}
\end{equation}
The functions $\psi _{\bot \beta }\left( \mathbf{\rho }\right) $ and $%
\varphi _{\beta \varepsilon }\left( z\right) $ satisfy the following
conditions: 
\begin{equation}
\psi _{\bot \beta }\left( \left| \mathbf{\rho }\right| =a\right) =0;
\end{equation}
\begin{equation}
\left. \frac{d\varphi _{\beta \varepsilon }\left( z\right) }{dz}\right|
_{z_{0}+0}-\left. \frac{d\varphi _{\beta \varepsilon} \left( z\right) }{dz}
\right| _{z_{0}-0}=\frac{2m^{\ast }g}{\hslash ^{2}}\left| \psi _{\bot \beta
}\left( \mathbf{\rho }_{0}\right) \right| ^{2}\varphi _{\beta \varepsilon
}\left( z_{0}\right) ;  \label{phi_z0}
\end{equation}
\begin{equation}
\varphi _{\beta \varepsilon }^{inc}\left( z\right) \rightarrow e^{ikz},\quad 
\mathrm{for }\quad z\rightarrow -\infty ;  \label{phi_inf}
\end{equation}
\begin{equation}
\left. \frac{d\varphi _{\beta \varepsilon }\left( z\right) }{dz}\right|
_{+0}-\left. \frac{d\varphi _{\beta \varepsilon }\left( z\right) }{dz}
\right| _{-0}=\frac{2m^{\ast }U}{\hslash ^{2}}\varphi _{\beta \varepsilon
}\left( 0\right) ;  \label{Upsi_z0}
\end{equation}
where $k=\sqrt{2m^{\ast }\varepsilon }/\hbar$ and $\varphi _{\beta
\varepsilon }^{inc}\left( z\right) $ is the incident wave. Condition (\ref%
{phi_inf}) means that we consider a wave $\varphi _{\beta \varepsilon
}^{inc}\left( z\right) $ of unit amplitude, which moves from $-\infty $
towards the contact.

For the subsequent calculations we make the explicit choice for the shape of
the contact $a\left( z\right) =a_{0}\cosh \left( z/L\right) .$ The condition
of adiabaticity for this dependence $a\left( z\right) $ is $L\gg \lambda
_{F}.$ The solution of Eq.~(\ref{Schred_ro}) \ is given by 
\begin{equation}
\psi _{\bot \beta }\left( \rho ,\varphi ,z\right) = \frac{1}{\sqrt{\pi }
a\left( z\right) J_{m+1}\left( \gamma _{mn}\right) }J_{m}\left( \gamma _{mn} 
\frac{\rho }{a\left( z\right) }\right) e^{im\varphi };  \label{psi_beta}
\end{equation}
having eigenvalues 
\begin{equation}
\varepsilon _{\beta } =\frac{\hbar ^{2}\gamma _{mn}^{2}}{2m^{\ast
}a^{2}\left( z\right) },\quad n=0,1,2,...;\quad m=0,\pm 1,\pm 2,...\quad
\label{e_beta}
\end{equation}
Here, we use cylindrical coordinates $\mathbf{\rho }=\left( \rho ,\varphi
,z\right)$ and $\gamma _{mn}$ is the $n-th$ zero of Bessel function $J_{m}.$
The energy spectrum (\ref{e_beta}) describes the quantized energy levels
inside the constriction ($z<L$) and a quasi-continuous spectrum at $z\gg L$
(the distance between the levels $\Delta \varepsilon _{\beta }\rightarrow 0$
at $\left| z\right| \rightarrow \infty $).

First, we consider a contact without an explicit barrier $\left( U=0\right) $%
. A general solution for the longitudinal wave function $\varphi _{\beta
\varepsilon }\left( z\right) $ in Eq.~(\ref{Schred_z}) has the form, 
\begin{eqnarray}
\varphi _{\beta \varepsilon }\left( z\right) &=&A\left( 1-\xi ^{2}\right)
^{- \frac{ikL}{2}}F\left( -ikL-s;-ikL+s+1;-ikL+s;\frac{1-\xi }{2}\right) +
\label{phi_hyper} \\
&&B\left( 1-\xi ^{2}\right) ^{-\frac{ikL}{2}}\left( \frac{1-\xi }{2}\right)
^{ikL}F\left( -s;s+1;1+ikL;\frac{1-\xi }{2}\right) ;  \notag
\end{eqnarray}
where $F\left( a,b,c;\xi \right) $ is the hypergeometric function, $\xi
=\tanh \left( \frac{z}{L}\right) ,$ and $s=\frac{1}{2}\left( -1+i\sqrt{%
\left( 2L\gamma _{mn}/a_{0}\right) ^{2}-1}\right) $ . The constants $A$ and $%
B$ can be found from the conditions (\ref{phi_z0}) and (\ref{phi_inf}). By
using the asymptotic form of hypergeometric function at $z>z_{0}\gg L$ and
in the limit of a small electron-impurity interaction constant $m^{\ast
}gk_{F}/ \hbar ^{2}\ll 1,$ we find 
\begin{equation}
\varphi _{\beta \varepsilon }\left( z\right) =t_{\beta }\left[ 1+\frac{%
gm^{\ast }}{ik\hbar ^{2}}\left| \psi _{\bot \beta }\left( \mathbf{\rho }%
_{0};z_{0}\right) \right| ^{2}\left( 1+r_{\beta }e^{ikz_{0}}\right) \right]
e^{ikz};  \label{phiz>z0}
\end{equation}
where $r_{\beta }$ and $t_{\beta }$ are the amplitudes of the reflected and
transmitted waves far from the contact 
\begin{eqnarray}
r_{\beta } &=&\frac{\Gamma \left( -ikL-s\right) \Gamma \left(
-ikL+s+1\right) \Gamma \left( ikL\right) }{\Gamma \left( -s\right) \Gamma
\left( s+1\right) \Gamma \left( -ikL\right) }; \\
t_{\beta } &=&\frac{\Gamma \left( -ikL-s\right) \Gamma \left(
-ikL+s+1\right) }{\Gamma \left( -ikL\right) \Gamma \left( 1-ikL\right) }
\end{eqnarray}

The expression for the current takes the form: 
\begin{equation}
I\left( V\right) =\frac{e}{\pi \sqrt{2m^{\ast }}}\sum_{\beta
}\int\limits_{z\gg L}d\mathbf{S}\int \frac{d\varepsilon }{\sqrt{\varepsilon }
}\ \mathrm{Im}\left( \psi _{\alpha }^{\ast }\nabla \psi _{\alpha }\right) %
\left[ n_{F}\left( \varepsilon \right) -n_{F}\left( \varepsilon +eV\right) %
\right] .  \label{current_ballist}
\end{equation}
Substituting the wave function $\psi _{\mathbf{\alpha }}\left( \mathbf{\rho }%
,z\right) $ from (\ref{psi_separ}) into Eq.~(\ref{current_ballist}), with $%
\psi _{\bot \beta }\left( \mathbf{\rho },z\right) $ given by (\ref{psi_beta}%
), and $\varphi _{\beta \varepsilon }\left( z\right) $ by (\ref{phiz>z0}),
and carrying out the integration over $\rho \leqslant a\left( z\right) $ at $%
z>z_{0}$, we find the total current trough the contact for $kL\gg 1,$ and $%
L\gg a_{0}$, 
\begin{gather}
I\left( V\right) =\frac{2e}{h}\int d\varepsilon \sum_{\beta }T_{\beta }\left[
n_{F}\left( \varepsilon \right) -n_{F}\left( \varepsilon +eV\right) \right]
\times \\
\left\{ 1+\frac{gm^{\ast }}{\hbar ^{2}k}\left| \psi _{\bot \beta }\left( 
\mathbf{\rho }_{0}\right) \right| ^{2}\left| r_{\beta }\right| \cos \left(
2kz_{0}+\varphi _{\beta }\right) \right\} .  \notag
\end{gather}
Here, 
\begin{equation}
T_{\beta }\left( \varepsilon \right) =\frac{1}{1+\exp \left[ 2\pi L\left(
k_{\beta }-k\right) \right] };  \label{T(e)}
\end{equation}
\begin{equation}
\left| r_{\beta }\left( \varepsilon \right) \right| =\frac{1}{\sqrt{\exp %
\left[ 2\pi L\left( k-k_{\beta }\right) \right] +1}};
\end{equation}
\begin{gather}
\varphi _{\beta }\left( \varepsilon \right) =\left( k+k_{\beta }\right)
L\left\{ 1-\ln \left( \left( k+k_{\beta }\right) L\right) \right\} \\
-\left( k-k_{\beta }\right) L\psi \left( \frac{1}{2}\right) -2kL(1-\ln kL); 
\notag
\end{gather}
$k_{\beta }=$ $\sqrt{2m^{\ast }\varepsilon _{\beta }\left( 0\right) }/\hbar $
is the quantized momentum of the transverse electron motion at the centre of
the contact; $\psi \left( \frac{1}{2}\right) =\Gamma ^{\prime }\left( \frac{1%
}{2}\right) /\Gamma \left( \frac{1}{2}\right) \thickapprox -1.96.$

The conductance $G\left( V\right) =dI/dV$, in the low temperature limit, is
given by 
\begin{equation}
G\left( V\right) =G_{0}\sum_{\beta }T_{\beta }\left( \varepsilon _{F}\right)
\left\{ 1+\frac{gm^{\ast }}{\hbar ^{2}k_{F}}\left| \psi _{\bot \beta }\left( 
\mathbf{\rho }_{0}\right) \right| ^{2}\left| r_{\beta }\right| \cos \left( 2%
\widetilde{k}_{F}z_{0}+\varphi _{\beta }\right) \right\} ,
\label{Cond_ballistic}
\end{equation}
where $G_{0}=2e^{2}/h$ is the quantum of conductance, and $\widetilde{k}_{F}=%
\sqrt{k_{F}^{2}+2meV/\hbar ^{2}}$. All energy dependent functions are taken
at $\varepsilon =\varepsilon _{F}$. We also used the condition $%
eV/\varepsilon _{F}\ll 1$.

The transmission coefficient is exponentially small $T_{\beta }\left(
\varepsilon \right) \sim \exp \left[ -2\pi L\left( k_{\beta }-k\right) %
\right] $ for $\varepsilon \ll \varepsilon _{\beta }\left( 0\right) ,$ while 
$T_{\beta }\left( \varepsilon \right) \rightarrow 1$ above this energy, $%
\varepsilon \gg \varepsilon _{\beta }\left( 0\right) .$ For $\left|
\varepsilon -\varepsilon _{\beta }\left( 0\right) \right| \ll \varepsilon
_{\beta }\left( 0\right) $ Eq.~(\ref{T(e)}) agrees with the formula for the
transmission coefficient that can be obtained in such case for an arbitrary
dependence $a\left( z\right) ,$ by the using an expansion near the point of
minimum cross section, $a\left( 0\right) $ (see, e.g., Ref. \cite{Term}).
For very long constrictions, $L\rightarrow \infty ,$ Eq.~(\ref{T(e)})
transforms to a step function $\Theta \left( \varepsilon -\varepsilon
_{\beta }\left( 0\right) \right) .$ \ For large $L$ the electrons are
strongly reflected by the contact when $k\simeq k_{\beta }.$ Hence, for
observation of conductance oscillations in an adiabatic ballistic
constriction the contact diameter should be chosen in such a way that $%
\varepsilon _{F}\gtrsim \varepsilon _{\beta }\left( 0\right) ,$ i.e., not
very far from the middle of a conductance step.

In the case $U\neq 0$ the boundary condition (\ref{Upsi_z0}) must be taken
into account. At $k\gg k_{\beta }$ reflection due to the shape of the
contact is negligibly small, as discuss above, and the conductance $G\left(
V\right) $ is described by the same equation (\ref{Cond_ballistic}), but
with 
\begin{equation}
T_{\beta }\left( \varepsilon \right) =\frac{1}{1+\left( m^{\ast }U/\hbar
^{2}k_{\beta }^{\prime }\right) ^{2}},
\end{equation}
\begin{equation}
\left| r_{\beta }\left( \varepsilon \right) \right| =\frac{m^{\ast }U/\hbar
^{2}k_{\beta }^{\prime }}{\sqrt{1+\left( m^{\ast }U/\hbar ^{2}k_{\beta
}^{\prime }\right) ^{2}}},
\end{equation}
\begin{equation}
\varphi _{\beta }\left( \varepsilon \right) =\arcsin \left( \frac{1}{\sqrt{%
1+\left( m^{\ast }U/\hbar ^{2}k_{\beta }^{\prime }\right) ^{2}}}\right) ,
\end{equation}
where $k_{\beta }^{\prime }=\sqrt{k^{2}-k_{\beta }^{2}}$.

\begin{figure}[t]
\includegraphics[width=0.7\linewidth]{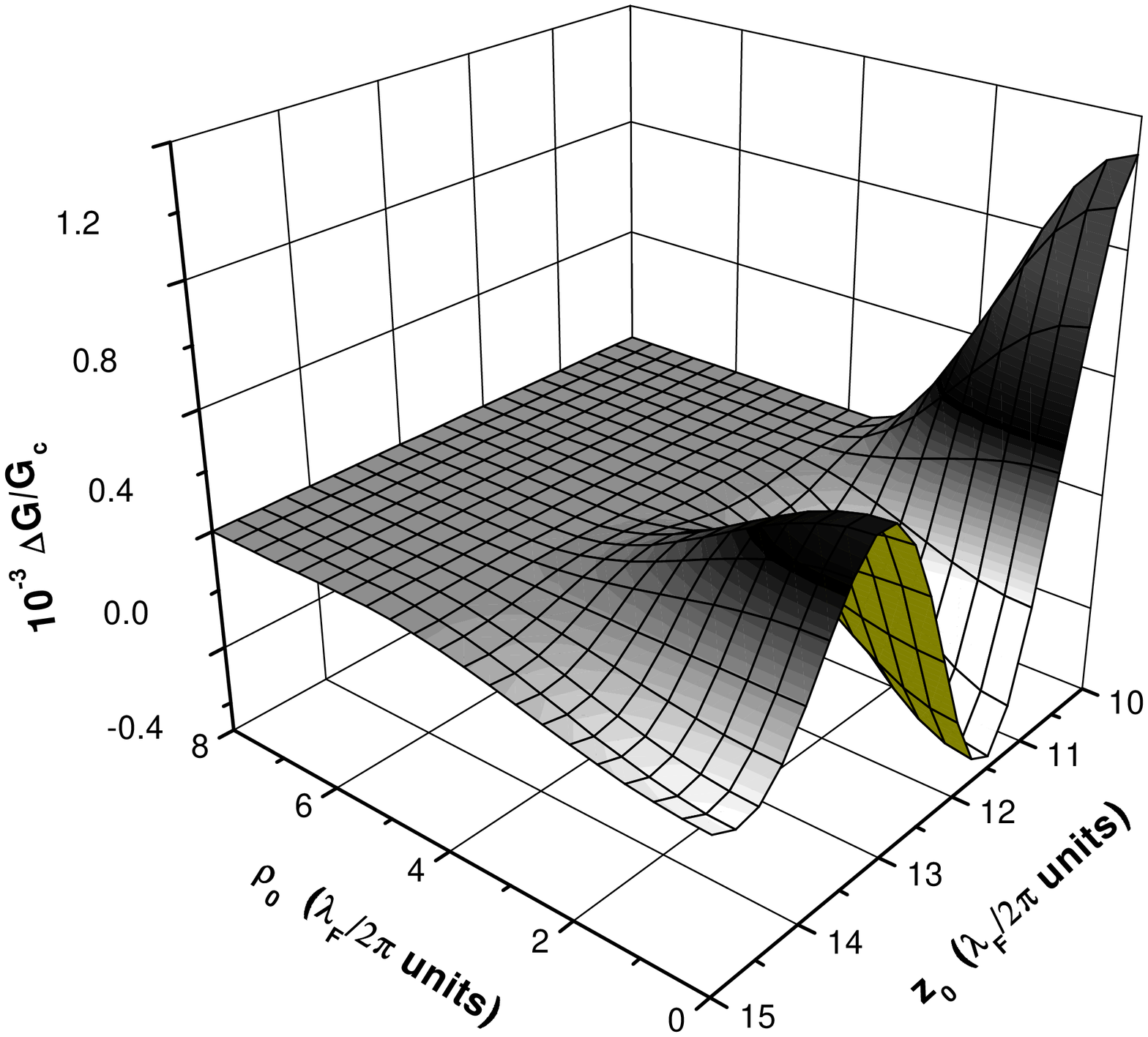}
\caption{Dependence of the oscillatory part of normalized conductance $%
\Delta G\left( 0\right) /G_{c}\left( 0\right) $ of the ballistic point
contact with the position coordinates of the defect $\left( \protect\rho %
_{0},z_{0}\right) $. \texttt{\ }}
\label{fig.G_ballistic_vs_Z0_and_rho0}
\end{figure}

\begin{figure}[t]
\includegraphics[width=0.7\linewidth]{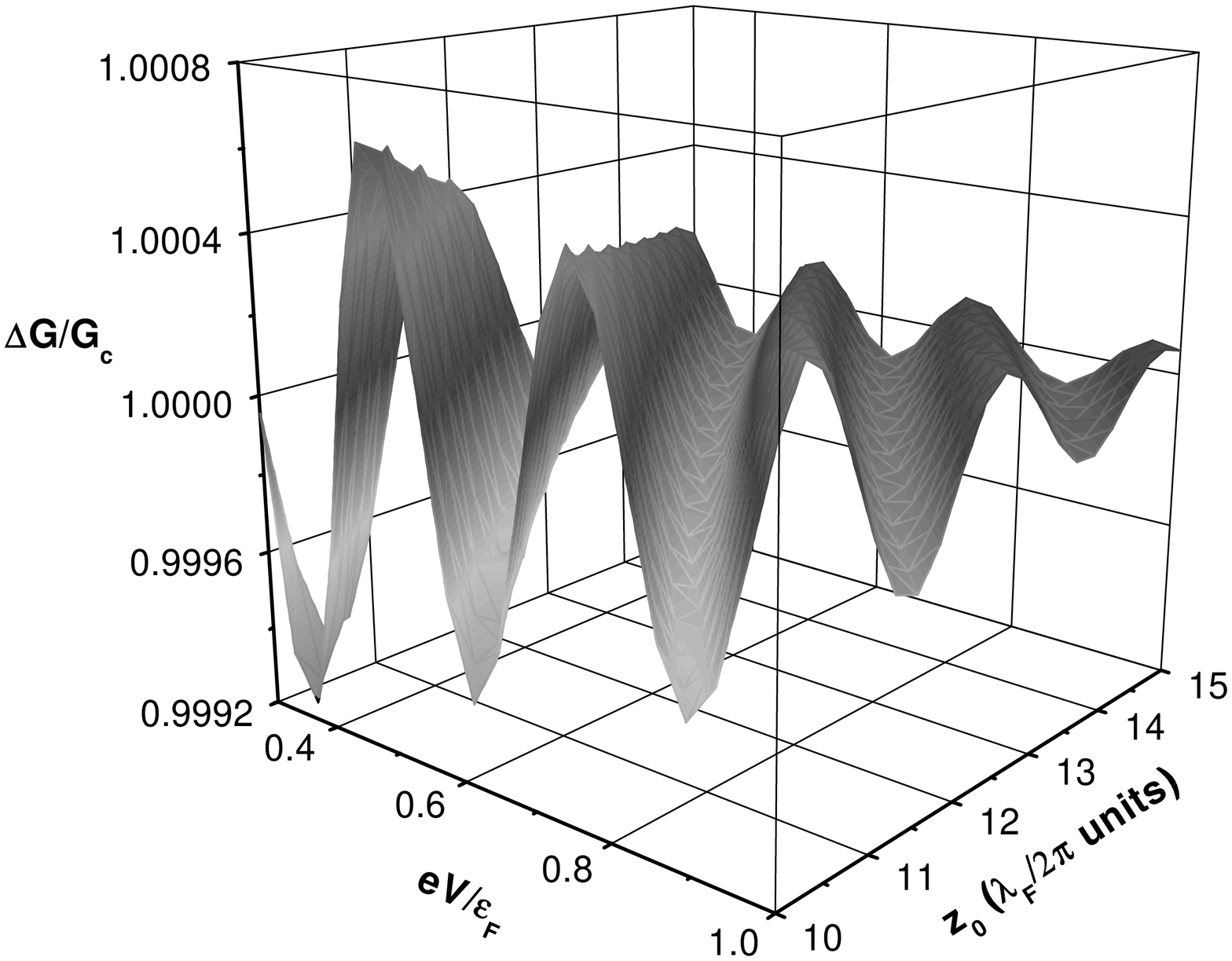}
\caption{Voltage dependence of the oscillatory part of normalized
conductance $\Delta G\left( V\right) /G_{c}\left( V\right) $ for an
adiabatic ballistic point contact for $\protect\rho _{0}=0$ as a function of
the depth $z_{0}$ of the defect under metal surface. \texttt{\ }}
\label{fig.G_ballistic_vs_Z0_and_V}
\end{figure}

Figure~\ref{fig.G_ballistic_vs_Z0_and_rho0} shows the dependence of the
oscillatory part of conductance $\Delta G(V)=G\left( V\right) -G_{c}\left(
V\right) $, Eq.~(\ref{Cond_ballistic}), on the position of the defect at low
voltage, $V\rightarrow 0$, for a contact without barrier $\left( U=0\right) $%
. Here, $G_{c}=G_{0}\sum_{\beta }T_{\beta }$ is the conductance in the
absence of a defect ($g=0$). Figure~\ref{fig.G_ballistic_vs_Z0_and_V} shows
the dependence of the $\Delta G(V)$ on applied bias voltage for a defect
sitting on the axis of the contact $\left( \rho _{0}=0\right) ,$ and as a
function of the distance $z_{0}$ from the contact center. In creating the
plots of Figs.~\ref{fig.G_ballistic_vs_Z0_and_rho0} and \ref%
{fig.G_ballistic_vs_Z0_and_V} we have used dimensionless parameter $2\pi
m^{\ast }gk_{F}/\hbar ^{2}=0.5$ , $2\pi a_{0}=2.405\lambda _{F}$ and $2\pi
L=10\lambda _{F},$ corresponding to a contact having one allowed quantum
conductance mode.

\section{\protect\bigskip Discussion.}

The presence of an elastic scattering center located inside the bulk, either
in the vicinity of a tunnel contact in an STM configuration or in one of the
banks of a ballistic point contact, has been shown to cause oscillatory
fluctuations in the conductance of the junction. For small contact radii ($%
a\ll \lambda _{F}$), these oscillations result solely from interference of
electron waves that are directly transmitted on the one hand, and electrons
that are both backscattered by the defect and again reflected by the contact
on the other. What now follows is a discussion whether this effect can be
employed experimentally for three dimensional mapping of subsurface
impurities.

In the case of a tunnel contact, the oscillatory part of the conductance can
be expressed by

\begin{equation}
\frac{G^{osc}(V)}{G_{c}}\propto \frac{z_{0}^{2}\lambda _{F}^{2}}{\gamma ^{4}}%
\sin 2\widetilde{k}_{F}\gamma ,\quad \widetilde{k}_{F}\gamma \gg 1;
\end{equation}
where $\widetilde{k}_{F}=\sqrt{k_{F}^{2}+2meV/\hbar ^{2}}$ is the wave
vector of electrons that are passing through the orifice and $z_{0}$ is the
depth of the defect under the surface; $\gamma $ is the distance between the
contact and the defect. Comparing this to the results found for a ballistic
contact, where

\begin{equation}
\frac{G^{osc}(V)}{G_{c}}\propto \frac{\lambda _{F}^{2}}{a^{2}\left(
z_{0}\right) }\cos \left( 2\widetilde{k}_{F}z_{0}+\varphi _{\beta }\right)
\end{equation}
(here $\varphi _{\beta }$ is the phase the electron acquires after
reflection by the contact), we see that although both oscillations have
similar arguments, the expression for the ballistic case has an extra phase $%
\varphi _{\beta }$ which depends nonlinearly on the wave vector $k$, making
the signal hard to identify. Secondly, the adiabatic condition, being an
essential assumption in the ballistic model, cannot be readily achieved
experimentally.

Therefore, choosing the tunnel contact for experimental application seems
most sensible. In that situation we can expect the information in the
conductance signal about a defect's whereabouts to be twofold: the amplitude
will decrease with growing distance $\gamma $, whereas the frequency of the
oscillation is expected to increase upon enlarging the distance from contact
to defect. The actual experiment would consist of sensitively measuring $%
\frac{dI}{dV}(V)$ curves on a tight grid of $\mathbf{\rho }$ coordinates.
The lateral positions of defects could then be identified as the centers of
radially symmetric patterns in this signal. Next, the depth of an impurity
should be derived from the period of the oscillation in the $\frac{dI}{dV}%
(V) $ curve at $\mathbf{\rho _{0}}$.

Assuming the numerical parameter $2\pi m^{\ast }gk_{F}/\hbar ^{2}=0.5$
introduced in Sec. III (which can be shown to be applicable for hard wall
scatterers with atomic radius) and choosing the orifice to be located
exactly above the defect ($\gamma =z_{0}$), the amplitude of the oscillation
is expected to be $10^{-1}G_{c}$ for $z_{0}$=3~nm (with $k_{F}=10^{10}$~$%
\text{m}^{-1}$).

Note that the choosing value of interaction constant is rather large. We use
the such value of the parameter to show more clear the investigated effects
\ in illustrations. For real value of parameter $g\approx 10^{-35} erg\cdot
cm^{3}$, which can be estimated from an electron effective scattering cross
section $\sim 1$\AA $^{2}$, the relative amplitude of oscillations is $%
10^{-2}\div 10^{-3}G_{c}$.

Comparing this to previous STS experiments \cite{Stipe}, where
signal-to-noise ratios of $5\cdot 10^{-4}$ (at 1~nA, 400~Hz sample
frequency) have been achieved, we should be able to measure defects located
more than 10 atomic layers under the surface.

As the period of the oscillation becomes longer for small $z_{0}$, the
minimum discernable depth will be determined by the maximum voltage that can
be applied over the junction. For example, 30~mV is sufficient for probing a
quarter of a conductance oscillation caused by a defect at 1~nm depth. The
increase of the noise level inherent to measuring at elevated voltages will
not pose a problem, as the amplitude of the signal is much higher for small
depths.

Finally, the anisotropy of the electronic structure will have to be taken
into account. Materials with an almost spherical Fermi surface such as Al or
Au, realizing the condition of a free electron gas, are expected to be most
suitable. Furthermore, deviations of spherical symmetry might be used as a
secondary proof for the effectiveness of the method, i.e. in the case of
Au(111), where the `necks' in the Fermi surface should cause a defect to be
invisible when probed exactly from above.

This research was supported partly by the program \textquotedblright
Nanosystems, nanomaterials and nanotechnology\textquotedblright\ of National
Academy of Sciences of Ukraine. Ye. S. Avotina wishes to acknowledge the
INTAS grant for Young Scientists.


\bigskip

\begin{thebibliography}{99}
\bibitem{Ebert} Ph. Ebert, M. Heinrich, M. Simon, C. Domke, K. Urban, C. K.
Shih, M. B. Webb, and M. G. Lagally, Phys. Rev. B \textbf{53}, 4580 (1996).

\bibitem{Heinze} S. Heinze, R. Abt, S. Bl\"{u}gel, G. Gilarowski, and H.
Niehus, Phys. Rev. Lett. \textbf{83}, 4808 (1999).

\bibitem{Schmid} M. Schmid, W. Hebenstreit, P. Varga, and S. Crampin, Phys.
Rev. Lett. \textbf{76}, 2298 (1996).

\bibitem{Hongbin} Hongbin Yu, C. S. Jiang, Ph. Ebert, and C. K. Shih, Appl.
Phys. Lett. \textbf{81}, 2005 (2002).

\bibitem{Crommie} M. F. Crommie, C. P. Lutz, and D. M. Eigler, Nature 
\textbf{363}, 524 (1993); ibid., Science \textbf{262}, 218 (1993).

\bibitem{Crampin} S. Crampin, J. Phys.: Condens. Matter \textbf{6}, L613
(1994).

\bibitem{Untiedt} C. Untiedt, G. Rubio Bollinger, S. Vieira, and N. Agra{%
\"{\i}}t, Phys. Rev. B, \textbf{62}, 9962 (2000).

\bibitem{Ludoph} B. Ludoph and J. M. van Ruitenbeek, Phys. Rev. B, \textbf{61%
}, 2273 (2000).

\bibitem{Kempen} A. Halbritter, Sz. Csonka, G. Mih{\'{a}}ly, O. I.
Shklyarevskii, S. Speller, and H. van Kempen, Phys. Rev. B, \textbf{69},
121411 (2004).

\bibitem{ISh} I. F. Itskovich and R. I. Shekhter, Fiz. Nizk. Temp., \textbf{%
11}, 373 (1985) [Sov. J. Low Temp. Phys., \textbf{11}, 202 (1985)].

\bibitem{Namir} A. Namiranian, Yu. A. Kolesnichenko, and A. N. Omelyanchouk,
Phys. Rev. B, \textbf{61}, 16796 (2000).

\bibitem{Avotina} Ye. S. Avotina, and Yu. A. Kolesnichenko, Fiz. Nizk.
Temp., \textbf{30, } 209 (2004) [J. Low Temp. Phys., \textbf{30}, 153
(2004)].

\bibitem{Avotina1} Ye. S. Avotina, A. Namiranian, and Yu. A. Kolesnichenko,
Phys. Rev. B, \textbf{70}, 075908 (2004).

\bibitem{KMO} I. O. Kulik, Yu. N. Mitsai, and A. N. Omelyanchouk, Zh. Exp.
Teor. Fiz., \textbf{63}, 1051 (1974).

\bibitem{Glazman} L. I. Glazman, G. B. Lesovik, D. E. Khmel'nitskii, and R.
I. Shekhter, \textit{JETP Lett.} \textbf{48}, 238 (1988).

\bibitem{3D} E. N. Bogachek, A. M. Zagoskin, and I. O. Kulik, \textit{Sov.
J. Low Temp. Phys.} \textbf{16}, 796 (1990).

\bibitem{LandauQM} L. D. Landau and E. M. Lifshits, Quantum Mechanics,
Pergamon, Oxford (1977).

\bibitem{Term} E. N. Bogachek, A. G. Scherbakov, and Uzi Landman, Phys. Rev.
B, \textbf{54}, R11094 (1996).

\bibitem{Stipe} B. C. Stipe, M. A. Rezaei, and W. Ho, Rev. Sci. Instr. 
\textbf{70}, 137 (1999).
\end{thebibliography}
\end{document}